\documentclass[prx,amsmath,amssymb,notitlepage,superscriptaddress,longbibliography]{revtex4-2}

\pdfoutput=1

\usepackage{graphicx}
\usepackage{color}

\def\bc{\begin{center}}
\def\ec{\end{center}}

\def\beq{\begin{equation}}
\def\eeq{\end{equation}}

\def\br{{\bf r}}

\def\br{{\bf r}}
\def\bq{{\bf q}}
\def\bk{{\bf k}}

\def\cK{{\cal K}}

\usepackage[papersize={8.5in,11in}]{geometry}

\usepackage{color}
\definecolor{darkblue}{rgb}{0.,0.,0.4}
\definecolor{darkred}{rgb}{0.5,0.,0.}
\definecolor{BlueViolet}{RGB}{138,43,226}
\definecolor{SkyBlue}{RGB}{30,144,255}
\definecolor{DarkGreen}{RGB}{0,100,0}
\usepackage[pdftex,colorlinks=true,linkcolor=darkblue,citecolor=blue,urlcolor=darkred]{hyperref}

\begin{document}

\title{
Robust quantum transport at particle-hole symmetry
}

\begin{abstract}
We study quantum transport in disordered systems with particle-hole symmetric Hamiltonians. 
The particle-hole symmetry is spontaneously broken after averaging with respect to disorder, and the resulting massless mode is treated in a random-phase representation of the invariant measure of the symmetry-group. We compute the resulting fermionic functional integral of the average two-particle Green's function in a perturbation theory around the diffusive limit. The results up to two-loop order show that the corrections vanish, indicating that the diffusive quantum transport is robust. On the other hand, the diffusion coefficient depends strongly on the particle-hole symmetric Hamiltonian we choose to study. This reveals a connection between the underlying microscopic theory and the classical long-scale metallic behaviour of these systems.   
\end{abstract}

\author{Ipsita Mandal}
\affiliation{Institute of Nuclear Physics, Polish Academy of Sciences, 31-342 Krak\'{o}w, Poland}
\affiliation{Department of Physics, Stockholm University, AlbaNova University Center,
106 91 Stockholm, Sweden}

\author{Klaus Ziegler}
\affiliation{
Institut f\"ur Physik, Universit\"at Augsburg,
D-86135 Augsburg, Germany}

\date{\today}

\maketitle

\section{Introduction}

Recent studies have found that transport in multi-band semimetals is substantially different from conventional transport based on the classical Boltzmann theory. This is due to the particle-hole (PH) symmetry, which is realized, at least approximately, in multi-band systems when the Fermi energy is between two neighboring bands. A typical example is 
the Dirac node of graphene. The PH symmetry of this two-band model leads to characteristic quantum effects, such as spontaneous particle-hole pair creation on arbitrarily small energy scales, which are only limited by the band width of the material. This is accompanied by strong fluctuations (also known as ``zitterbewegung''), which causes a finite dc conductivity even in the absence of disorder. Although graphene is a two-dimensional (2d) material, where fluctuation effects are strong, these quantum fluctuations may also play a crucial role in higher dimensions. Based on this idea, there has been a lot of progress to compute transport properties in systems like the 3d Weyl semimetals 
\cite{PhysRevB.33.3257,PhysRevB.33.3263,PhysRevB.83.205101,PhysRevLett.106.056401,
PhysRevB.84.235126,PhysRevLett.107.186806,PhysRevLett.108.140405,PhysRevLett.108.046602,
PhysRevB.85.195320,PhysRevB.86.115208,PhysRevB.85.035103,PhysRevB.87.235306,PhysRevB.89.014205,
PhysRevLett.110.236803,PhysRevB.87.155123,PhysRevLett.112.016402,PhysRevB.89.054202,
PhysRevB.90.241112,PhysRevLett.113.026602,PhysRevLett.114.166601,Ziegler2016,pixley2021rare}.

The fundamental quantity for the study of quantum transport is the transition probability $P_{\br\br'}$, for a particle to move from site $\br'$ to a site $\br$ on a $d$-dimensional lattice. This classical quantity can be linked to a quantum model through the average two-particle Green's function (A2PGF) of a particle of energy $E$, given by
\beq
K_{\br\br'}=\frac{1}{\pi} 
\left \langle G_{\br\br'}(E+i \,\epsilon)
\,G_{\br'\br}(E-i\, \epsilon)\right \rangle_d
\ \ \
( \text{where } \epsilon>0)\,.
\label{2pgf}
\eeq
Here $H$ is a random Hamiltonian,
$G_{\br\br'}(E+i \,\epsilon)=\left \langle\br'|(H-E-i \,\epsilon)^{-1}|\br\right \rangle$
is the one-particle Green's function, and $\left \langle \ldots \right \rangle_d$ denotes averaging with respect to disorder-induced randomness. Then the transition probability is given by
$P_{\br\br'}=K_{\br\br'}/\sum \limits_{\br'}
K_{\br\br'}$. The time-dependent transition probability
is obtained through the Fourier transformation $E\to t$.
The transport properties in the metallic regime can be understood by computing the diffusion coefficient $D$, which can be obtained from $K_{\br\br'}$ as~\cite{THOULESS197493}
\beq
D=\lim_{\epsilon\to0}\epsilon^2\sum \limits_{\br\in \Lambda} r^2  \,K_{0 \br}\,,
\label{diffusion_coeff00}
\eeq
on a lattice $\Lambda$. The corresponding dc conductivity is related to $D$ via the Einstein relation.

We can assume that the form of $G$ results from a self-energy approximation in interacting many-body systems. Then $\epsilon$, as the imaginary part of the self-energy, depends on the
frequency $\omega$ and the Fermi energy $E_F$. With this, we can bridge the microscopic quantum modeling and the more qualitative hydrodynamic description of transport due to long-lived modes in quantum systems~\cite{PhysRevB.76.144502,LUCAS2015239,PhysRevD.89.066018,hartnoll2018holographic,ips_andy}. In particular, using this formalism, we can analyze the effect of a vanishing $\epsilon \sim \omega^s$, where $s$ is a positive rational number.

In this paper, we will focus on diffusion in systems with PH symmetry, in the
presence of disorder. We assume that the disorder also obeys the PH symmetry. Although systems without
PH symmetry can also be treated by the subsequently discussed method (cf. Refs.~~\cite{PhysRevB.86.155450,Ziegler2013}), we shall focus here only on the PH-symmetric case for simplicity.
Taking into account the fact that averaging over a random distribution
of disorder leads to spontaneous PH symmetry-breaking, we employ a field theory representation to study this effect. Although the PH symmetry is discrete, an underlying global symmetry of the field theory is continuous \cite{PhysRevB.79.195424}. Thus, there is a massless mode associated with the spontaneous PH symmetry-breaking that leads to long-range correlations. These correlations are the origin of diffusion. They can be calculated from the massless mode, using the integration over the symmetry related saddle point manifold. This is known as the integration with respect to the invariant measure of the symmetry group, and is often approximated in a leading order gradient expansion, also known as the nonlinear sigma model approach \cite{WEGNER1987210}. 
Here we will consider the full invariant measure, which was shown to provide a simple expression for the  A2PGF in terms of a random-phase model \cite{Ziegler_2015}.
This is briefly summarized in Sec.~\ref{sect:rpa}.

The paper is organized as follows. The representation of the invariant measure by the random-phase model is briefly reviewed in Sec.~\ref{sect:rpa}. In Sec.~\ref{sect:fir}, we introduce a fermionic functional integral for the description of the A2PGF, and employ a perturbation expansion around the diffusive approximation. It is shown that one- and two-loop corrections vanish, indicating the robustness of
diffusion. In Sec.~\ref{sect:discussion} we study some examples of the microscopic Hamiltonians with PH symmetry, and the results reveal that the diffusion coefficients depend strongly on the details of these microscopic Hamiltonians.

\section{random-phase representation of the average two-particle Green's function}
\label{sect:rpa}

For simplicity, we are restricting the discussion to two bands, while an extension to $n$ bands is straightforward. 
Hence, our starting point is a two-band Hamiltonian $H$ with matrix elements $H_{\br j,\br'j'}$, where $\br,\br'$ are the co-ordinates on a $d$-dimensional lattice, and $j,j'$ are band indices. In this case, we can assign a Pauli matrix representation for the Hamiltonian as 
\beq
H_{\br j,\br' j'}=\sum \limits_{ a=1}^3 \mathcal{H}_{a;\br\br'}\, 
\sigma^a_{jj'}\, ,
\eeq
where $\sigma^a $ is a Pauli matrix, and $ \mathcal{H}_{a;\br\br'}$ a matrix element on the lattice. 
An instructive example is discussed in Appendix~\ref{app:example}.
The $\br$-independent PH transformation $S$ transforms the Hamiltonian $H$ as $S\,H\,S^{-1}=-H$, which implies that an eigenvector $\Psi_E$ with energy $E$ is related to the eigenvector $\Psi_{-E}$ with energy $-E$ by $\Psi_{-E}=S\,\Psi_E$. Thus, $S$ is a symmetry transformation for the Green's function $G(z)\to -S\,G(z) \,S^{-1}=G(-z)$ at $z=0$, which is exactly at the mirror-symmetric point between the two symmetric bands. Due to the poles of $G(z)$, this
symmetry-point must be treated with care. We can avoid the poles by choosing $z=i \,\epsilon$ ($\epsilon>0$). 
Then the difference, in the limit $\epsilon\to 0$, reads
\beq
\lim_{\epsilon\to0} \left [-S\,G(i \,\epsilon)\,S^{-1}-G(i \,\epsilon) \right ]
=\lim_{\epsilon\to0} \left [G(-i \,\epsilon)-G(i \,\epsilon) \right ] .
\label{spont.ssb}
\eeq
A nonzero result indicates a spontaneously-broken PH symmetry. For the diagonal elements of the Green's functions, the right-hand side is proportional to the density of states at $E=0$. This reflects the fact that a nonzero density of states at $E=0$ provides a sufficient condition for spontaneous PH symmetry-breaking.

The disorder averaged one-particle Green's function can be calculated within the self-consistent Born approximation (or the saddle-point approximation of a functional-integral representation) as
\beq
\left \langle H\pm i \,\epsilon\right \rangle_d  ^{-1}\approx 
\left [H_0\pm i \left (\epsilon+\eta \right) \right ]^{-1},
 \quad 
H_0=\left \langle H\right \rangle_d +\Sigma',
\label{scba}
\eeq
where 
$\Sigma'$ is the real part of the self-energy, and $\eta$ is its imaginary part.
The parameter $\eta$ provides a broadening of the average one-particle Green's function, and
also plays the role of an order parameter for spontaneous PH symmetry-breaking, since 
we get $2\,i\,\eta \left (H_0^2+\eta^2 \right )^{-1}$ for Eq.~(\ref{spont.ssb}).

Transport properties are determined by properties on large time and spatial scales.
In Ref.~\cite{Ziegler_2015}, it was shown that the long-range part of the A2PGF of Eq.~(\ref{2pgf}) can be obtained by reducing the disorder average to the integration with respect to the invariant measure of the saddle-point manifold. This means that in practice, we replace $H$ with the effective 
random-phase Hamiltonian
\beq
{\cal H}_R= U \,H_0 \, U^\dagger
\ ,\ \ \
U=\text{diag}(e^{i \,\alpha_{\br j}})\,,
\eeq
and then average with respect to the independently and identically distributed random-phases $\{\alpha_{\br j}\}$.
This gives us
\beq
K_{\br\br'}
\sim{\cal K}_{\br\br'}
=\frac{\left \langle adj_{{\bar \br}{\bar \br}'}
\,C\right \rangle_\alpha}{\left \langle  \det C\right \rangle_\alpha}
\ ,\ \
\left \langle \ldots \right \rangle_\alpha
=\frac{1}{2\pi}\int_0^{2\pi} \ldots 
\prod \limits_{\br,j}d\alpha_{\br j}\,,
\label{corr1a}
\eeq
with the random-phase matrix
\beq
C_{\br\br'}= 2\delta_{\br\br'}-\sum \limits_{j,j'}
e^{i \,\alpha_{\br j}} \,h_{\br j,\br'j'}
\sum \limits_{j'',\br''} h^\dagger_{\br'j',\br''j''}
\,e^{-i \,\alpha_{\br''j''}}
\ ,
\eeq
with
\beq 
h_{\br\br'}=\mathcal{I}_2 \,\delta_{\br\br'}
+ 2\,i \,\eta \left (H_0 - i \,{\bar\eta} \right)^{-1}_{\br\br'}
\,,\quad 
{\bar\eta}=\eta+\epsilon\,,
\label{def_h}
\eeq
and $adj_{{\bar \br}{\bar \br}'} C$ denoting the elements of the adjugate matrix.
Under a PH transformation, we obtain the Hermitian conjugation $S\,h\,S^{-1}=h^\dagger$, which implies that $C$ is real and symmetric. The quantity $U\,h\,U^\dagger$ represents the effective one-particle Green's function of the generic system, while $C$ is the corresponding effective two-particle propagator.

Although $C_{\br\br'}$ is still a random matrix, the A2PGF in Eq.~(\ref{corr1a}) is much simpler to treat than the A2PGF in Eq.~(\ref{2pgf}), because the phase integration is not plagued by poles of the integrand. Nevertheless, this does not mean that the theory becomes simple. For instance, the long-range behaviour of the A2PGF is based on the zero modes of $C$, which exist for any realization of the random-phases due to the relation
\beq
h\,h^\dagger
={\bf 1}-4 \,\epsilon \left(1-\epsilon \right ){\bar\eta}
\left( H_0^2+{\bar\eta}^2 \right)^{-1}
\ .
\label{unitary0}
\eeq
This implies that the constant mode $\Psi_0$ with vanishing wavevector $\bk=0$ obeys
\beq
\sum \limits_{\br'}C_{\br\br'}\Psi_0\sim c \,\epsilon \,\Psi_0
\ ,
\eeq
i.e., $\Psi_0$ is always a zero-energy eigenmode of the effective two-particle propagator in the limit $\epsilon\to 0$.

In order to evaluate $\cK$, or the diffusion coefficient $D$, we can employ two
different methods. The first is based on a graphical representation, while the second involves a fermionic functional integral representation. While the former was described and
discussed in Ref.~\cite{Ziegler_2015}, we will focus on the latter in this paper.

\subsection{General properties of ${\tilde \cK}_\bq$}

Before we start with the specific calculations, it is useful to mention an important 
connection between the two-particle and the one-particle Green's function (``Ward Identity''), which takes the form
\beq
{\tilde K}_{\bq=0}=\frac{\pi}{\epsilon}\left \langle\rho(0)\right \rangle_d
\ ,
\eeq
after averaging. Here, $\left \langle\rho(0)\right \rangle_d$ is the disorder average of the density of states
at energy $E=0$, which is typically nonzero and finite. Although we do not
have a proof, this relation should also hold for $\cK$ due to $K\sim \cK$ on large scales. Therefore, 
${\tilde \cK}_{\bq=0}\sim \text{const.} \,\epsilon^{-1}$, which will be confirmed in the subsequent calculation.
The second derivative with respect to $q_\mu$ at $\bq=0$ is
\beq
-\partial_{q_\mu}^2{\tilde K}_\bq\Big|_{\bq=0}\equiv -{\tilde K}_0''
=D\epsilon^{-\beta}
\ ,
\eeq
with $\beta=2$ for diffusion. This is in agreement with Eq.~(\ref{diffusion_coeff00}).
Higher-order derivatives of ${\tilde \cK}_\bq$ are also of interest, since
\beq
\sum \limits_{\br} ({r_\mu})^{2n}\,K_{0\br}
=(-1)^n \,{\tilde K}^{(2n)}_0
\eeq
describe higher moments of spatial fluctuations.


\section{Functional integral representation}
\label{sect:fir}

\begin{figure}[t]
\begin{center}
\includegraphics[width=6.5cm,height=4cm]{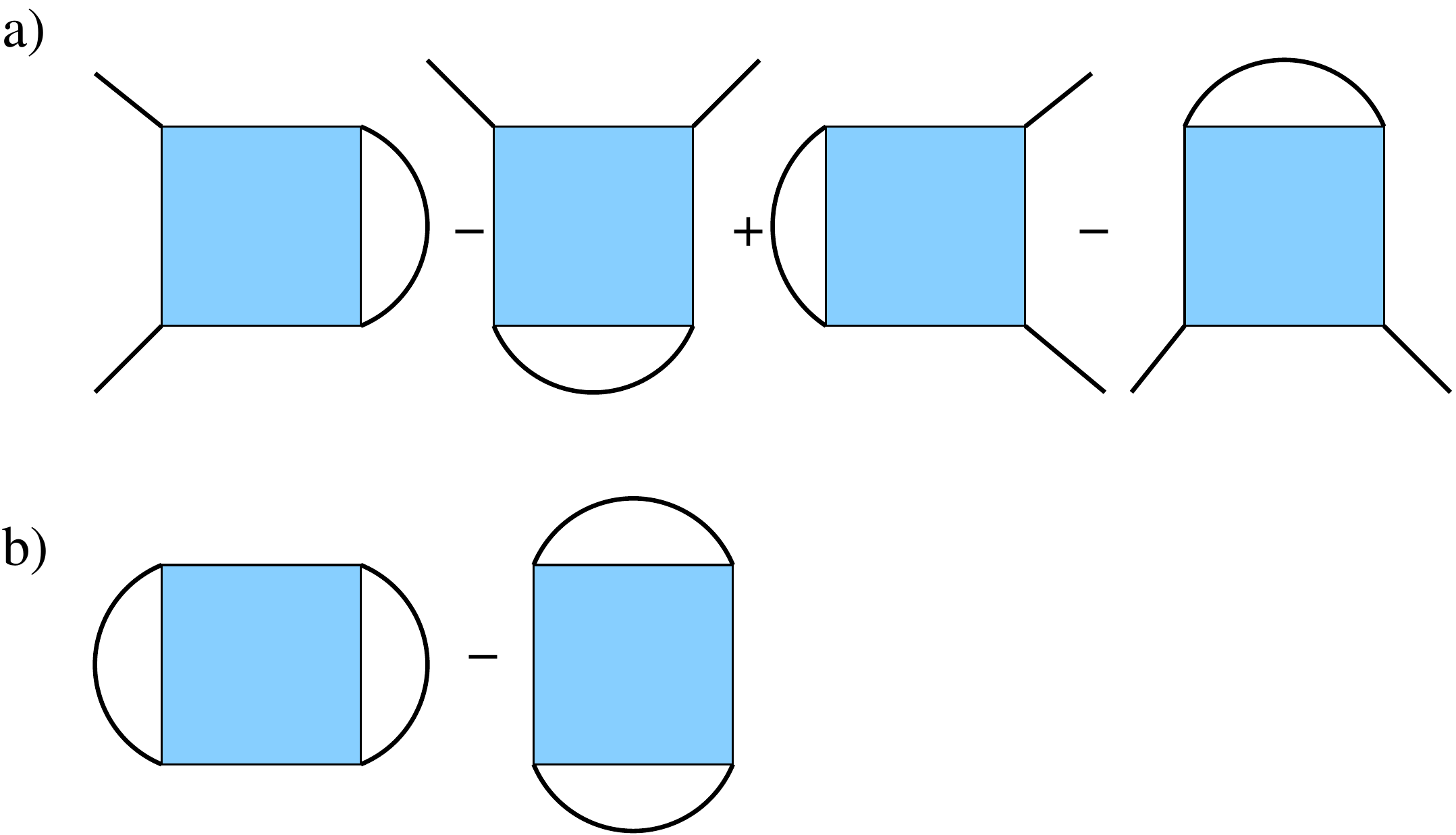}
\caption{\label{fig:feynman}
First order perturbation theory of $\cK$ in Eq.~(\ref{prop2}) around the diffusion propagator $G$ for the (a) numerator,  and (b) the denominator. These terms cancel each other due to the symmetry of the blue square vertex. 
}
\end{center}
\end{figure}

The averaged determinant in Eq.~(\ref{corr1a}) can be expressed as a fermionic (Grassmann) functional integral
\cite{zbMATH01727741}
\beq
\left \langle\det C\right \rangle_\alpha = \int_\Psi \left \langle\exp(-\Psi\cdot C \,\bar{\Psi})\right \rangle_\alpha ,
\eeq
which implies that
\beq
\cK_{\br\br'}=\frac{\int_\Psi \bar{\Psi}_\br\Psi_{\br'}\left \langle\exp(-\Psi\cdot C \,\bar{\Psi})\right \rangle_\alpha}
{\int_\Psi \left \langle\exp(-\Psi\cdot C \,\bar{\Psi})\right \rangle_\alpha} \,.
\label{prop2}
\eeq
In terms of a perturbation theory around $\bar{C}=\left \langle C\right \rangle_\alpha$,
with
\beq
\left \langle C_{\br\br'}\right \rangle_\alpha 
=2 \, \delta_{\br\br'}-\sum \limits_{j,j'=1,2}
h_{\br j,\br'j'} \, h^\dagger_{\br'j',\br j}
=2\,\delta_{\br\br'}-Tr_2 \left (h_{\br\br'} \,h^\dagger_{\br'\br} \right ),
\label{av_c00}
\eeq
we have
\beq
\left \langle e^{-\Psi\, C \,\bar{\Psi}} \right \rangle_\alpha
=\left \langle
e^{-\Psi \, \left (\bar{C}-C' \right )\bar{\Psi}}
\right \rangle_\alpha
=\left [1+\left \langle \left (\Psi\, C'\,\bar{\Psi} \right )^2 \right \rangle_\alpha/2 
+ \ldots
\right ] e^{-\Psi\, \bar{C} \,\bar{\Psi}}\ .
\eeq
Using this, we obtain
\beq
\int_\Psi \bar{\Psi}_\br \,\Psi_{\br'}\left \langle
e^{-\Psi\, C \,\bar{\Psi}} \right \rangle_\alpha
=\int_\Psi\bar{\Psi}_\br \, \Psi_{\br'}[1+\left \langle(\Psi\, C'\,\bar{\Psi})^2\right \rangle_\alpha/2 +...]
e^{-\Psi\, \bar{C}  \,\bar{\Psi}},
\eeq
and
\beq
\int_\Psi \left \langle
e^{-\Psi\, C \,\bar{\Psi}}\right \rangle_\alpha
=\int_\Psi \left [1+\left \langle(\Psi\, C'\,\bar{\Psi})^2\right \rangle_\alpha/2 +
\ldots \right ]
e^{-\Psi\, \bar{C} \,\bar{\Psi}}\ .
\eeq
After normalizing both the expressions by $Z_0=\det\left \langle C\right \rangle_\alpha$, we can represent the result
graphically by two-loop graphs as depicted in Fig.~\ref{fig:feynman} (neglecting higher order terms indicated by $ \ldots $). The square in Fig.~\ref{fig:feynman} represents the vertex
\beq
V_{\br_1\br_2\br_3\br_4}
\equiv \left \langle C'_{\br_1\br_2} \, C'_{\br_3\br_4}\right \rangle_\alpha
=Tr_2 \left ( h_{\br_1\br_2} \,h^\dagger_{\br_2\br_3}
\,h_{\br_3\br_4} \,h^\dagger_{\br_4\br_1} \right )
-\delta_{\br_1\br_3} \sum \limits_j
\left (h_{\br_1\br_2}\,h^\dagger_{\br_2\br_1}\right )_{jj} 
\left (h_{\br_1\br_4}\,h^\dagger_{\br_4\br_1} \right )_{jj},
\label{vertex}
\eeq
and the thick lines are the unperturbed propagator $\bar{C}^{-1}$. It turns out that the two-loop corrections cancel each other. This is a consequence of the anti-commuting property of the fermion field. 
The details of the calculation are given in Appendix~\ref{app:details}.
Thus, in the Fourier space only the unperturbed propagator 
\beq
{\tilde \cK}_\bq =\frac{1}{2-\int_\bk 
Tr_2 \left ({\tilde h}_\bk \,{\tilde h}^\dagger_{\bk-\bq} \right )}
\label{diff_prop1}
\eeq
survives in this approximation. Here $\int_\bk$ denotes the normalized integral with respect to the 
$d$-dimensional sphere with radius $\lambda$.  
The denominator of ${\tilde \cK}_\bq$ can be expanded in powers of $\bq$. This leads to ${\tilde \cK}_\bq\sim 1/(A\,\epsilon+B\,q^2)$, provided the expansion exists and the system is isotropic. Due to the mirror-symmetric dispersion $(E_\bk,-E_\bk)$ of the PH-symmetric averaged Hamiltonian $H_0$,
we get
\beq
A=8\,\eta\int_\bk\frac{1}{E_\bk^2+\bar\eta^2}
\, ,\ \ 
B=\frac{1}{2}\int_\bk Tr_2\left (\tilde{h}_\bk \,\partial_{k_\mu}\partial_{k_\mu} \tilde{h}^\dagger_\bk \right ) .
\label{diff_prop2}
\eeq
Thus, ${\tilde \cK}_\bq$ is a diffusion propagator with the diffusion coefficient $D=B/A$.
This perturbative result clearly indicates that diffusion is quite robust for a PH-symmetric Hamiltonian. 
The robustness of diffusion in terms of a perturbation theory was also observed for the special case of 2d Dirac fermions \cite{PhysRevB.55.10661,PhysRevB.79.195424}.

\section{Discussions}
\label{sect:discussion}

For a better understanding of the result in Eqs. (\ref{diff_prop1}) and (\ref{diff_prop2}), 
we will consider a simple example of a two-band Hamiltonian. It is defined in the Fourier space with the dispersion $E_\bk$ as
\beq
{\tilde h}_\bk
=\sigma_0+\frac{2\,i\,\eta}{E_\bk^2+\bar{\eta}^2}
\begin{pmatrix}
E_\bk+i \,{\bar\eta} & 0 \\
0 & -E_\bk+i\,{\bar\eta} \\
\end{pmatrix} ,
\eeq
which can also be written as
\beq
\tilde{h}_\bk= \begin{pmatrix}
\kappa_\bk & 0 \\
0 & \kappa_\bk^*\\
\end{pmatrix} ,\ \ \
\kappa_\bk=\frac{E_\bk^2-\bar{\eta}^2
+2\,\epsilon\,\bar{\eta}+2 \,i\,\eta\, E_\bk}{E_\bk^2+\bar{\eta}^2} \,.
\eeq
In the limit $\epsilon\to0$, we get a unimodular function
\beq
\kappa_\bk\to\frac{(E_\bk+i\,\eta)^2}{E_\bk^2+\eta^2}=e^{i\,\phi_\bk}
\ , \ \
\phi_\bk=\text{arg}\left [(E_\bk^2+i\,\eta)^2) \right].
\label{unimodular1}
\eeq
The Fourier transform of $\bar{C}$ takes the form
\beq
\tilde{C}_\bq=\int_\bk \left [2-Tr_2
\left ({\tilde h}_\bk \,{\tilde h}^\dagger_{\bk-\bq}\right ) \right ]
=\int_\bk \left (2-\kappa_\bk^*  \, \kappa_{\bk-\bq}
-\kappa_\bk\, \kappa_{\bk-\bq}^*  \right ) ,
\label{int_C}
\eeq
which gives the expression $\tilde{C}_\bq\sim A\epsilon+B\,q^2$ (as shown in Sec.~\ref{sect:fir}) for small $\bq$. Here $A$ is defined as the integral in Eq.~(\ref{diff_prop2}) and
\beq
B=4\,\eta^2\int_\bk
\frac{\left (\partial_{k_\mu} E_\bk \right )\left (\partial_{k_\mu} E_\bk \right )}
{\left (E_\bk^2+\eta^2 \right )^2} \ .
\eeq
Our model has two independent parameters, the effective disorder strength $\eta$ and the momentum cut-off $\lambda$, where $1/\lambda$ defines the shortest wavelength. Moreover, the dimensionality $d$ of the $\bk$-integration plays a crucial role. Although we do not expect that the qualitative behaviour of diffusion is much affected by the short-distance regime, the diffusion coefficient $D=B/A$ might depend on it.
In order to study this effect and its relation to different dispersions $E_\bk$, we calculate it for two characteristic examples, namely $E_\bk=E_s\, k^s$ with $s=1,2$. The expressions for $A$ and $B$
imply that the energy coefficient $E_s$ can be absorbed in the scaling of $\eta$ and $\epsilon$. Therefore, we implicitly assume subsequently that these parameters are scaled as $\eta\to\eta/E_s$ and $\epsilon\to\epsilon/E_s$.

The integral in Eq.~(\ref{int_C}) can be calculated for $\epsilon\sim0$, we obtain the results:
\begin{enumerate}
\item $E_\bk=k$, $d=2$:
\beq
\tilde{C}_\bq\sim \frac{4\,\eta} {3\,\lambda^4}
\left [ 
-\eta \,q^2+
6\,\eta^2\,\epsilon-3\,\eta\,\lambda^2+ 3\,\lambda^2 
\left \lbrace
\frac{\eta  \left (q^2+2 \,\eta^2 \right )
\ln \left (1+ \frac {q\left (q+\sqrt{q^2+4 \,\eta^2} \right ) } {2 \,\eta^2} \right )}
{q \,\sqrt{4 \,\eta^2+q^2}}-4\,\epsilon  \ln(\eta/\lambda)
\right \rbrace \right ],
\eeq
and
\beq
\tilde{C}_\bq \vert_{\mathbf q \sim 0}\sim\frac{4 \,\eta} {3\,\lambda}
\left [
\left \lbrace \frac{6 \,\eta^2} {\lambda^2}
-12\ln(\eta/\lambda)  \right \rbrace \frac{\epsilon}{\lambda} +
\frac{\lambda^2-\eta^2}{\eta} \,q^2
\right ].
\eeq

\item
$E_\bk=k$, $d=3$:
\beq
\tilde{C}_\bq\sim \frac{3\,\eta}{2\,\pi\,\lambda^3\,q}
\left [ \left (32\,\lambda\,\epsilon -2\,\pi\,\eta^2 \right ) q
+\pi\,\eta \left (4\,\eta^2+3\,q^2 \right )
{\rm arctan} \left ( \frac{q } {2\,\eta} \right )   \right ]
\eeq
and
\beq
\tilde{C}_\bq \vert_{\mathbf q \sim 0} 
\sim \frac{3\,\eta} {2\,\pi\,\lambda}
\left(\frac{32}{\lambda}\epsilon +\frac{4\pi}{3\,\lambda^2}q^2\right).
\eeq
This describes diffusion with a diffusion coefficient 
$D=\lambda^2
\left ({\tilde\eta}^{-1}-{\tilde\eta} \right )/
\left (6\,{\tilde\eta}^2-12\ln {\tilde\eta} \right )$
with ${\tilde\eta}= \eta/\lambda$ in $d=2$ and
$D=\frac {\pi } {24  \,\lambda}$ in $d=3$.

\item
$E_\bk=k^2$:
\beq
\tilde{C}_\bq\sim
\begin{cases}
4 \, \lambda^{-2} \left [\epsilon  \left (\pi-2\eta/\lambda^2 \right )
+q^2/\lambda^2 \right ] 
& \text{for } d=2\,, \\
\lambda^{-3}\sqrt{2\,\eta} \left (6\epsilon+\eta \,q^2 \right ) & \text{ for } d=3\,,
\end{cases}
\eeq
for $\epsilon\sim0$ and $q\sim0$.
This describes diffusion with diffusion coefficients $D=1/(\pi\lambda^2-2\, \eta)$ ($d=2$)
and $D=\eta/6$ ($d=3$). It is remarkable that $D$ vanishes with $\eta\to 0$ only in $d=3$,
but not in $d=2$. 

\end{enumerate}

The results for the diffusion coefficients $D$ are summarized in Table \ref{table1}.
They clearly indicate that these diffusion coefficients depend strongly on the dispersion of $H_0$. A vanishing order parameter $\eta$ indicates a transition from the metallic phase to another phase, typically to an insulating phase. The results in Table \ref{table1} reveal that the properties of such a transition from the metallic side depend strongly on the details of the model and the dimensionality of the system. Although we focus on the metallic phase here, these properties might be interesting and deserve a further analysis to identify the strong influences of the PH-symmetric Hamiltonians. 

\begin{table}
\begin{tabular}{| l | l | l |}
\hline
 & $d=2$ & $d=3$ \\ \hline
$E_\bk=k$ & 
$
\frac{\lambda^2\left ({\tilde\eta}^{-1}-{\tilde\eta} \right )}
{6 \, {\tilde\eta}^2-12  \ln \tilde \eta }$ 
& $\frac{\pi } {24\,\lambda} $\\
$E_\bk=k^2$ & 
$ \frac{1}
{\lambda^2  \left ( \pi-2\eta / \lambda^2 \right)} $ 
& $\eta/6$ \\
Dirac fermions & $-\frac{1}
{4\, \lambda \,{\tilde\eta} \ln{\tilde\eta} }$ & $-$ \\
\hline
\end{tabular}
\caption{The diffusion coefficients for the dispersions $E_\bk $, and for dimensionalities $d=2,3$, with ${\tilde\eta}=\eta/\lambda$. The result for 2d Dirac fermions is from Ref.~\cite{PhysRevB.79.195424}.
}
\label{table1}
\end{table}

In conclusion, we have found that diffusion (i.e., metallic behaviour) is very robust if the
model Hamiltonian obeys the PH transformation property $H\to S\,H\,S^{-1}=-H$, as the one-loop and two-loop corrections vanish. The reason behind this is
the spontaneous PH symmetry-breaking, which is associated with a spontaneous breaking of a continuous symmetry, which creates a massless mode. However, the diffusion coefficient $D$ is very sensitive to the spectral properties of $H$ and the dimensionality of the underlying space. This connection between the underlying microscopic details of the model and the classical diffusion coefficient is an advantage of our approach, in comparison to more heuristic approaches (e.g., the Mori-Zwanzig memory matrix formalism \cite{Forster-HFBSCF,Freire-AP_2017,hartnoll2018holographic,ips-freire1,ips-freire2}). In particular, it will be interesting to apply this formalism to Luttinger semimetals, where methods like Kubo formula and memory matrix fail unless the PH symmetry is broken by unequal band masses \cite{rahul-sid,ips-rahul,ips-freire1,ips-freire2}. Finally, it will be worthwhile to see if our formalism can be used to compute transport properties in non-Fermi liquids having a critical Fermi surface \cite{ips-uv-ir1,ips-uv-ir2,ips-subir,ips-c2,ips-fflo,ips-nfl-u1}.

\section{Acknowledgments} 
KZ gratefully acknowledges the support by the Julian Schwinger Foundation.

\bibliography{ref_weyl}

\begin{thebibliography}{47}%
\makeatletter
\providecommand \@ifxundefined [1]{%
 \@ifx{#1\undefined}
}%
\providecommand \@ifnum [1]{%
 \ifnum #1\expandafter \@firstoftwo
 \else \expandafter \@secondoftwo
 \fi
}%
\providecommand \@ifx [1]{%
 \ifx #1\expandafter \@firstoftwo
 \else \expandafter \@secondoftwo
 \fi
}%
\providecommand \natexlab [1]{#1}%
\providecommand \enquote  [1]{``#1''}%
\providecommand \bibnamefont  [1]{#1}%
\providecommand \bibfnamefont [1]{#1}%
\providecommand \citenamefont [1]{#1}%
\providecommand \href@noop [0]{\@secondoftwo}%
\providecommand \href [0]{\begingroup \@sanitize@url \@href}%
\providecommand \@href[1]{\@@startlink{#1}\@@href}%
\providecommand \@@href[1]{\endgroup#1\@@endlink}%
\providecommand \@sanitize@url [0]{\catcode `\\12\catcode `\$12\catcode
  `\&12\catcode `\#12\catcode `\^12\catcode `\_12\catcode `\%12\relax}%
\providecommand \@@startlink[1]{}%
\providecommand \@@endlink[0]{}%
\providecommand \url  [0]{\begingroup\@sanitize@url \@url }%
\providecommand \@url [1]{\endgroup\@href {#1}{\urlprefix }}%
\providecommand \urlprefix  [0]{URL }%
\providecommand \Eprint [0]{\href }%
\providecommand \doibase [0]{https://doi.org/}%
\providecommand \selectlanguage [0]{\@gobble}%
\providecommand \bibinfo  [0]{\@secondoftwo}%
\providecommand \bibfield  [0]{\@secondoftwo}%
\providecommand \translation [1]{[#1]}%
\providecommand \BibitemOpen [0]{}%
\providecommand \bibitemStop [0]{}%
\providecommand \bibitemNoStop [0]{.\EOS\space}%
\providecommand \EOS [0]{\spacefactor3000\relax}%
\providecommand \BibitemShut  [1]{\csname bibitem#1\endcsname}%
\let\auto@bib@innerbib\@empty
\bibitem [{\citenamefont {Fradkin}(1986{\natexlab{a}})}]{PhysRevB.33.3257}%
  \BibitemOpen
  \bibfield  {author} {\bibinfo {author} {\bibfnamefont {E.}~\bibnamefont
  {Fradkin}},\ }\bibfield  {title} {\bibinfo {title} {Critical behavior of
  disordered degenerate semiconductors. i. models, symmetries, and formalism},\
  }\href {https://doi.org/10.1103/PhysRevB.33.3257} {\bibfield  {journal}
  {\bibinfo  {journal} {Phys. Rev. B}\ }\textbf {\bibinfo {volume} {33}},\
  \bibinfo {pages} {3257} (\bibinfo {year} {1986}{\natexlab{a}})}\BibitemShut
  {NoStop}%
\bibitem [{\citenamefont {Fradkin}(1986{\natexlab{b}})}]{PhysRevB.33.3263}%
  \BibitemOpen
  \bibfield  {author} {\bibinfo {author} {\bibfnamefont {E.}~\bibnamefont
  {Fradkin}},\ }\bibfield  {title} {\bibinfo {title} {Critical behavior of
  disordered degenerate semiconductors. ii. spectrum and transport properties
  in mean-field theory},\ }\href {https://doi.org/10.1103/PhysRevB.33.3263}
  {\bibfield  {journal} {\bibinfo  {journal} {Phys. Rev. B}\ }\textbf {\bibinfo
  {volume} {33}},\ \bibinfo {pages} {3263} (\bibinfo {year}
  {1986}{\natexlab{b}})}\BibitemShut {NoStop}%
\bibitem [{\citenamefont {Wan}\ \emph {et~al.}(2011)\citenamefont {Wan},
  \citenamefont {Turner}, \citenamefont {Vishwanath},\ and\ \citenamefont
  {Savrasov}}]{PhysRevB.83.205101}%
  \BibitemOpen
  \bibfield  {author} {\bibinfo {author} {\bibfnamefont {X.}~\bibnamefont
  {Wan}}, \bibinfo {author} {\bibfnamefont {A.~M.}\ \bibnamefont {Turner}},
  \bibinfo {author} {\bibfnamefont {A.}~\bibnamefont {Vishwanath}},\ and\
  \bibinfo {author} {\bibfnamefont {S.~Y.}\ \bibnamefont {Savrasov}},\
  }\bibfield  {title} {\bibinfo {title} {Topological semimetal and fermi-arc
  surface states in the electronic structure of pyrochlore iridates},\ }\href
  {https://doi.org/10.1103/PhysRevB.83.205101} {\bibfield  {journal} {\bibinfo
  {journal} {Phys. Rev. B}\ }\textbf {\bibinfo {volume} {83}},\ \bibinfo
  {pages} {205101} (\bibinfo {year} {2011})}\BibitemShut {NoStop}%
\bibitem [{\citenamefont {Smith}\ \emph {et~al.}(2011)\citenamefont {Smith},
  \citenamefont {Banerjee}, \citenamefont {Pardo},\ and\ \citenamefont
  {Pickett}}]{PhysRevLett.106.056401}%
  \BibitemOpen
  \bibfield  {author} {\bibinfo {author} {\bibfnamefont {J.~C.}\ \bibnamefont
  {Smith}}, \bibinfo {author} {\bibfnamefont {S.}~\bibnamefont {Banerjee}},
  \bibinfo {author} {\bibfnamefont {V.}~\bibnamefont {Pardo}},\ and\ \bibinfo
  {author} {\bibfnamefont {W.~E.}\ \bibnamefont {Pickett}},\ }\bibfield
  {title} {\bibinfo {title} {Dirac point degenerate with massive bands at a
  topological quantum critical point},\ }\href
  {https://doi.org/10.1103/PhysRevLett.106.056401} {\bibfield  {journal}
  {\bibinfo  {journal} {Phys. Rev. Lett.}\ }\textbf {\bibinfo {volume} {106}},\
  \bibinfo {pages} {056401} (\bibinfo {year} {2011})}\BibitemShut {NoStop}%
\bibitem [{\citenamefont {Burkov}\ \emph {et~al.}(2011)\citenamefont {Burkov},
  \citenamefont {Hook},\ and\ \citenamefont {Balents}}]{PhysRevB.84.235126}%
  \BibitemOpen
  \bibfield  {author} {\bibinfo {author} {\bibfnamefont {A.~A.}\ \bibnamefont
  {Burkov}}, \bibinfo {author} {\bibfnamefont {M.~D.}\ \bibnamefont {Hook}},\
  and\ \bibinfo {author} {\bibfnamefont {L.}~\bibnamefont {Balents}},\
  }\bibfield  {title} {\bibinfo {title} {Topological nodal semimetals},\ }\href
  {https://doi.org/10.1103/PhysRevB.84.235126} {\bibfield  {journal} {\bibinfo
  {journal} {Phys. Rev. B}\ }\textbf {\bibinfo {volume} {84}},\ \bibinfo
  {pages} {235126} (\bibinfo {year} {2011})}\BibitemShut {NoStop}%
\bibitem [{\citenamefont {Xu}\ \emph {et~al.}(2011)\citenamefont {Xu},
  \citenamefont {Weng}, \citenamefont {Wang}, \citenamefont {Dai},\ and\
  \citenamefont {Fang}}]{PhysRevLett.107.186806}%
  \BibitemOpen
  \bibfield  {author} {\bibinfo {author} {\bibfnamefont {G.}~\bibnamefont
  {Xu}}, \bibinfo {author} {\bibfnamefont {H.}~\bibnamefont {Weng}}, \bibinfo
  {author} {\bibfnamefont {Z.}~\bibnamefont {Wang}}, \bibinfo {author}
  {\bibfnamefont {X.}~\bibnamefont {Dai}},\ and\ \bibinfo {author}
  {\bibfnamefont {Z.}~\bibnamefont {Fang}},\ }\bibfield  {title} {\bibinfo
  {title} {Chern semimetal and the quantized anomalous hall effect in
  ${\mathrm{hgcr}}_{2}{\mathrm{se}}_{4}$},\ }\href
  {https://doi.org/10.1103/PhysRevLett.107.186806} {\bibfield  {journal}
  {\bibinfo  {journal} {Phys. Rev. Lett.}\ }\textbf {\bibinfo {volume} {107}},\
  \bibinfo {pages} {186806} (\bibinfo {year} {2011})}\BibitemShut {NoStop}%
\bibitem [{\citenamefont {Young}\ \emph {et~al.}(2012)\citenamefont {Young},
  \citenamefont {Zaheer}, \citenamefont {Teo}, \citenamefont {Kane},
  \citenamefont {Mele},\ and\ \citenamefont {Rappe}}]{PhysRevLett.108.140405}%
  \BibitemOpen
  \bibfield  {author} {\bibinfo {author} {\bibfnamefont {S.~M.}\ \bibnamefont
  {Young}}, \bibinfo {author} {\bibfnamefont {S.}~\bibnamefont {Zaheer}},
  \bibinfo {author} {\bibfnamefont {J.~C.~Y.}\ \bibnamefont {Teo}}, \bibinfo
  {author} {\bibfnamefont {C.~L.}\ \bibnamefont {Kane}}, \bibinfo {author}
  {\bibfnamefont {E.~J.}\ \bibnamefont {Mele}},\ and\ \bibinfo {author}
  {\bibfnamefont {A.~M.}\ \bibnamefont {Rappe}},\ }\bibfield  {title} {\bibinfo
  {title} {Dirac semimetal in three dimensions},\ }\href
  {https://doi.org/10.1103/PhysRevLett.108.140405} {\bibfield  {journal}
  {\bibinfo  {journal} {Phys. Rev. Lett.}\ }\textbf {\bibinfo {volume} {108}},\
  \bibinfo {pages} {140405} (\bibinfo {year} {2012})}\BibitemShut {NoStop}%
\bibitem [{\citenamefont {Hosur}\ \emph {et~al.}(2012)\citenamefont {Hosur},
  \citenamefont {Parameswaran},\ and\ \citenamefont
  {Vishwanath}}]{PhysRevLett.108.046602}%
  \BibitemOpen
  \bibfield  {author} {\bibinfo {author} {\bibfnamefont {P.}~\bibnamefont
  {Hosur}}, \bibinfo {author} {\bibfnamefont {S.~A.}\ \bibnamefont
  {Parameswaran}},\ and\ \bibinfo {author} {\bibfnamefont {A.}~\bibnamefont
  {Vishwanath}},\ }\bibfield  {title} {\bibinfo {title} {Charge transport in
  weyl semimetals},\ }\href {https://doi.org/10.1103/PhysRevLett.108.046602}
  {\bibfield  {journal} {\bibinfo  {journal} {Phys. Rev. Lett.}\ }\textbf
  {\bibinfo {volume} {108}},\ \bibinfo {pages} {046602} (\bibinfo {year}
  {2012})}\BibitemShut {NoStop}%
\bibitem [{\citenamefont {Wang}\ \emph {et~al.}(2012)\citenamefont {Wang},
  \citenamefont {Sun}, \citenamefont {Chen}, \citenamefont {Franchini},
  \citenamefont {Xu}, \citenamefont {Weng}, \citenamefont {Dai},\ and\
  \citenamefont {Fang}}]{PhysRevB.85.195320}%
  \BibitemOpen
  \bibfield  {author} {\bibinfo {author} {\bibfnamefont {Z.}~\bibnamefont
  {Wang}}, \bibinfo {author} {\bibfnamefont {Y.}~\bibnamefont {Sun}}, \bibinfo
  {author} {\bibfnamefont {X.-Q.}\ \bibnamefont {Chen}}, \bibinfo {author}
  {\bibfnamefont {C.}~\bibnamefont {Franchini}}, \bibinfo {author}
  {\bibfnamefont {G.}~\bibnamefont {Xu}}, \bibinfo {author} {\bibfnamefont
  {H.}~\bibnamefont {Weng}}, \bibinfo {author} {\bibfnamefont {X.}~\bibnamefont
  {Dai}},\ and\ \bibinfo {author} {\bibfnamefont {Z.}~\bibnamefont {Fang}},\
  }\bibfield  {title} {\bibinfo {title} {Dirac semimetal and topological phase
  transitions in ${A}_{3}$bi ($a=\text{Na}$, k, rb)},\ }\href
  {https://doi.org/10.1103/PhysRevB.85.195320} {\bibfield  {journal} {\bibinfo
  {journal} {Phys. Rev. B}\ }\textbf {\bibinfo {volume} {85}},\ \bibinfo
  {pages} {195320} (\bibinfo {year} {2012})}\BibitemShut {NoStop}%
\bibitem [{\citenamefont {Singh}\ \emph {et~al.}(2012)\citenamefont {Singh},
  \citenamefont {Sharma}, \citenamefont {Lin}, \citenamefont {Hasan},
  \citenamefont {Prasad},\ and\ \citenamefont {Bansil}}]{PhysRevB.86.115208}%
  \BibitemOpen
  \bibfield  {author} {\bibinfo {author} {\bibfnamefont {B.}~\bibnamefont
  {Singh}}, \bibinfo {author} {\bibfnamefont {A.}~\bibnamefont {Sharma}},
  \bibinfo {author} {\bibfnamefont {H.}~\bibnamefont {Lin}}, \bibinfo {author}
  {\bibfnamefont {M.~Z.}\ \bibnamefont {Hasan}}, \bibinfo {author}
  {\bibfnamefont {R.}~\bibnamefont {Prasad}},\ and\ \bibinfo {author}
  {\bibfnamefont {A.}~\bibnamefont {Bansil}},\ }\bibfield  {title} {\bibinfo
  {title} {Topological electronic structure and weyl semimetal in the
  tlbise${}_{2}$ class of semiconductors},\ }\href
  {https://doi.org/10.1103/PhysRevB.86.115208} {\bibfield  {journal} {\bibinfo
  {journal} {Phys. Rev. B}\ }\textbf {\bibinfo {volume} {86}},\ \bibinfo
  {pages} {115208} (\bibinfo {year} {2012})}\BibitemShut {NoStop}%
\bibitem [{\citenamefont {Hal\'asz}\ and\ \citenamefont
  {Balents}(2012)}]{PhysRevB.85.035103}%
  \BibitemOpen
  \bibfield  {author} {\bibinfo {author} {\bibfnamefont {G.~B.}\ \bibnamefont
  {Hal\'asz}}\ and\ \bibinfo {author} {\bibfnamefont {L.}~\bibnamefont
  {Balents}},\ }\bibfield  {title} {\bibinfo {title} {Time-reversal invariant
  realization of the weyl semimetal phase},\ }\href
  {https://doi.org/10.1103/PhysRevB.85.035103} {\bibfield  {journal} {\bibinfo
  {journal} {Phys. Rev. B}\ }\textbf {\bibinfo {volume} {85}},\ \bibinfo
  {pages} {035103} (\bibinfo {year} {2012})}\BibitemShut {NoStop}%
\bibitem [{\citenamefont {Liu}\ \emph {et~al.}(2013)\citenamefont {Liu},
  \citenamefont {Ye},\ and\ \citenamefont {Qi}}]{PhysRevB.87.235306}%
  \BibitemOpen
  \bibfield  {author} {\bibinfo {author} {\bibfnamefont {C.-X.}\ \bibnamefont
  {Liu}}, \bibinfo {author} {\bibfnamefont {P.}~\bibnamefont {Ye}},\ and\
  \bibinfo {author} {\bibfnamefont {X.-L.}\ \bibnamefont {Qi}},\ }\bibfield
  {title} {\bibinfo {title} {Chiral gauge field and axial anomaly in a weyl
  semimetal},\ }\href {https://doi.org/10.1103/PhysRevB.87.235306} {\bibfield
  {journal} {\bibinfo  {journal} {Phys. Rev. B}\ }\textbf {\bibinfo {volume}
  {87}},\ \bibinfo {pages} {235306} (\bibinfo {year} {2013})}\BibitemShut
  {NoStop}%
\bibitem [{\citenamefont {Biswas}\ and\ \citenamefont
  {Ryu}(2014)}]{PhysRevB.89.014205}%
  \BibitemOpen
  \bibfield  {author} {\bibinfo {author} {\bibfnamefont {R.~R.}\ \bibnamefont
  {Biswas}}\ and\ \bibinfo {author} {\bibfnamefont {S.}~\bibnamefont {Ryu}},\
  }\bibfield  {title} {\bibinfo {title} {Diffusive transport in weyl
  semimetals},\ }\href {https://doi.org/10.1103/PhysRevB.89.014205} {\bibfield
  {journal} {\bibinfo  {journal} {Phys. Rev. B}\ }\textbf {\bibinfo {volume}
  {89}},\ \bibinfo {pages} {014205} (\bibinfo {year} {2014})}\BibitemShut
  {NoStop}%
\bibitem [{\citenamefont {Kobayashi}\ \emph {et~al.}(2013)\citenamefont
  {Kobayashi}, \citenamefont {Ohtsuki},\ and\ \citenamefont
  {Imura}}]{PhysRevLett.110.236803}%
  \BibitemOpen
  \bibfield  {author} {\bibinfo {author} {\bibfnamefont {K.}~\bibnamefont
  {Kobayashi}}, \bibinfo {author} {\bibfnamefont {T.}~\bibnamefont {Ohtsuki}},\
  and\ \bibinfo {author} {\bibfnamefont {K.-I.}\ \bibnamefont {Imura}},\
  }\bibfield  {title} {\bibinfo {title} {Disordered weak and strong topological
  insulators},\ }\href {https://doi.org/10.1103/PhysRevLett.110.236803}
  {\bibfield  {journal} {\bibinfo  {journal} {Phys. Rev. Lett.}\ }\textbf
  {\bibinfo {volume} {110}},\ \bibinfo {pages} {236803} (\bibinfo {year}
  {2013})}\BibitemShut {NoStop}%
\bibitem [{\citenamefont {Huang}\ \emph {et~al.}(2013)\citenamefont {Huang},
  \citenamefont {Das}, \citenamefont {Balatsky},\ and\ \citenamefont
  {Arovas}}]{PhysRevB.87.155123}%
  \BibitemOpen
  \bibfield  {author} {\bibinfo {author} {\bibfnamefont {Z.}~\bibnamefont
  {Huang}}, \bibinfo {author} {\bibfnamefont {T.}~\bibnamefont {Das}}, \bibinfo
  {author} {\bibfnamefont {A.~V.}\ \bibnamefont {Balatsky}},\ and\ \bibinfo
  {author} {\bibfnamefont {D.~P.}\ \bibnamefont {Arovas}},\ }\bibfield  {title}
  {\bibinfo {title} {Stability of weyl metals under impurity scattering},\
  }\href {https://doi.org/10.1103/PhysRevB.87.155123} {\bibfield  {journal}
  {\bibinfo  {journal} {Phys. Rev. B}\ }\textbf {\bibinfo {volume} {87}},\
  \bibinfo {pages} {155123} (\bibinfo {year} {2013})}\BibitemShut {NoStop}%
\bibitem [{\citenamefont {Kobayashi}\ \emph {et~al.}(2014)\citenamefont
  {Kobayashi}, \citenamefont {Ohtsuki}, \citenamefont {Imura},\ and\
  \citenamefont {Herbut}}]{PhysRevLett.112.016402}%
  \BibitemOpen
  \bibfield  {author} {\bibinfo {author} {\bibfnamefont {K.}~\bibnamefont
  {Kobayashi}}, \bibinfo {author} {\bibfnamefont {T.}~\bibnamefont {Ohtsuki}},
  \bibinfo {author} {\bibfnamefont {K.-I.}\ \bibnamefont {Imura}},\ and\
  \bibinfo {author} {\bibfnamefont {I.~F.}\ \bibnamefont {Herbut}},\ }\bibfield
   {title} {\bibinfo {title} {Density of states scaling at the semimetal to
  metal transition in three dimensional topological insulators},\ }\href
  {https://doi.org/10.1103/PhysRevLett.112.016402} {\bibfield  {journal}
  {\bibinfo  {journal} {Phys. Rev. Lett.}\ }\textbf {\bibinfo {volume} {112}},\
  \bibinfo {pages} {016402} (\bibinfo {year} {2014})}\BibitemShut {NoStop}%
\bibitem [{\citenamefont {Ominato}\ and\ \citenamefont
  {Koshino}(2014)}]{PhysRevB.89.054202}%
  \BibitemOpen
  \bibfield  {author} {\bibinfo {author} {\bibfnamefont {Y.}~\bibnamefont
  {Ominato}}\ and\ \bibinfo {author} {\bibfnamefont {M.}~\bibnamefont
  {Koshino}},\ }\bibfield  {title} {\bibinfo {title} {Quantum transport in a
  three-dimensional weyl electron system},\ }\href
  {https://doi.org/10.1103/PhysRevB.89.054202} {\bibfield  {journal} {\bibinfo
  {journal} {Phys. Rev. B}\ }\textbf {\bibinfo {volume} {89}},\ \bibinfo
  {pages} {054202} (\bibinfo {year} {2014})}\BibitemShut {NoStop}%
\bibitem [{\citenamefont {Roy}\ and\ \citenamefont
  {Das~Sarma}(2014)}]{PhysRevB.90.241112}%
  \BibitemOpen
  \bibfield  {author} {\bibinfo {author} {\bibfnamefont {B.}~\bibnamefont
  {Roy}}\ and\ \bibinfo {author} {\bibfnamefont {S.}~\bibnamefont
  {Das~Sarma}},\ }\bibfield  {title} {\bibinfo {title} {Diffusive quantum
  criticality in three-dimensional disordered dirac semimetals},\ }\href
  {https://doi.org/10.1103/PhysRevB.90.241112} {\bibfield  {journal} {\bibinfo
  {journal} {Phys. Rev. B}\ }\textbf {\bibinfo {volume} {90}},\ \bibinfo
  {pages} {241112} (\bibinfo {year} {2014})}\BibitemShut {NoStop}%
\bibitem [{\citenamefont {Sbierski}\ \emph {et~al.}(2014)\citenamefont
  {Sbierski}, \citenamefont {Pohl}, \citenamefont {Bergholtz},\ and\
  \citenamefont {Brouwer}}]{PhysRevLett.113.026602}%
  \BibitemOpen
  \bibfield  {author} {\bibinfo {author} {\bibfnamefont {B.}~\bibnamefont
  {Sbierski}}, \bibinfo {author} {\bibfnamefont {G.}~\bibnamefont {Pohl}},
  \bibinfo {author} {\bibfnamefont {E.~J.}\ \bibnamefont {Bergholtz}},\ and\
  \bibinfo {author} {\bibfnamefont {P.~W.}\ \bibnamefont {Brouwer}},\
  }\bibfield  {title} {\bibinfo {title} {Quantum transport of disordered weyl
  semimetals at the nodal point},\ }\href
  {https://doi.org/10.1103/PhysRevLett.113.026602} {\bibfield  {journal}
  {\bibinfo  {journal} {Phys. Rev. Lett.}\ }\textbf {\bibinfo {volume} {113}},\
  \bibinfo {pages} {026602} (\bibinfo {year} {2014})}\BibitemShut {NoStop}%
\bibitem [{\citenamefont {Syzranov}\ \emph {et~al.}(2015)\citenamefont
  {Syzranov}, \citenamefont {Radzihovsky},\ and\ \citenamefont
  {Gurarie}}]{PhysRevLett.114.166601}%
  \BibitemOpen
  \bibfield  {author} {\bibinfo {author} {\bibfnamefont {S.~V.}\ \bibnamefont
  {Syzranov}}, \bibinfo {author} {\bibfnamefont {L.}~\bibnamefont
  {Radzihovsky}},\ and\ \bibinfo {author} {\bibfnamefont {V.}~\bibnamefont
  {Gurarie}},\ }\bibfield  {title} {\bibinfo {title} {Critical transport in
  weakly disordered semiconductors and semimetals},\ }\href
  {https://doi.org/10.1103/PhysRevLett.114.166601} {\bibfield  {journal}
  {\bibinfo  {journal} {Phys. Rev. Lett.}\ }\textbf {\bibinfo {volume} {114}},\
  \bibinfo {pages} {166601} (\bibinfo {year} {2015})}\BibitemShut {NoStop}%
\bibitem [{\citenamefont {Ziegler}(2016)}]{Ziegler2016}%
  \BibitemOpen
  \bibfield  {author} {\bibinfo {author} {\bibfnamefont {K.}~\bibnamefont
  {Ziegler}},\ }\bibfield  {title} {\bibinfo {title} {Quantum transport in 3d
  weyl semimetals: Is there a metal-insulator transition?},\ }\href
  {https://doi.org/10.1140/epjb/e2016-70454-2} {\bibfield  {journal} {\bibinfo
  {journal} {The European Physical Journal B}\ }\textbf {\bibinfo {volume}
  {89}},\ \bibinfo {pages} {268} (\bibinfo {year} {2016})}\BibitemShut
  {NoStop}%
\bibitem [{\citenamefont {Pixley}\ and\ \citenamefont
  {Wilson}(2021)}]{pixley2021rare}%
  \BibitemOpen
  \bibfield  {author} {\bibinfo {author} {\bibfnamefont {J.~H.}\ \bibnamefont
  {Pixley}}\ and\ \bibinfo {author} {\bibfnamefont {J.~H.}\ \bibnamefont
  {Wilson}},\ }\href@noop {} {\bibinfo {title} {Rare regions and avoided
  quantum criticality in disordered weyl semimetals and superconductors}}
  (\bibinfo {year} {2021}),\ \Eprint {https://arxiv.org/abs/2102.02822}
  {arXiv:2102.02822 [cond-mat.dis-nn]} \BibitemShut {NoStop}%
\bibitem [{\citenamefont {Thouless}(1974)}]{THOULESS197493}%
  \BibitemOpen
  \bibfield  {author} {\bibinfo {author} {\bibfnamefont {D.}~\bibnamefont
  {Thouless}},\ }\bibfield  {title} {\bibinfo {title} {Electrons in disordered
  systems and the theory of localization},\ }\href
  {https://doi.org/https://doi.org/10.1016/0370-1573(74)90029-5} {\bibfield
  {journal} {\bibinfo  {journal} {Physics Reports}\ }\textbf {\bibinfo {volume}
  {13}},\ \bibinfo {pages} {93} (\bibinfo {year} {1974})}\BibitemShut {NoStop}%
\bibitem [{\citenamefont {Hartnoll}\ \emph {et~al.}(2007)\citenamefont
  {Hartnoll}, \citenamefont {Kovtun}, \citenamefont {M\"uller},\ and\
  \citenamefont {Sachdev}}]{PhysRevB.76.144502}%
  \BibitemOpen
  \bibfield  {author} {\bibinfo {author} {\bibfnamefont {S.~A.}\ \bibnamefont
  {Hartnoll}}, \bibinfo {author} {\bibfnamefont {P.~K.}\ \bibnamefont
  {Kovtun}}, \bibinfo {author} {\bibfnamefont {M.}~\bibnamefont {M\"uller}},\
  and\ \bibinfo {author} {\bibfnamefont {S.}~\bibnamefont {Sachdev}},\
  }\bibfield  {title} {\bibinfo {title} {Theory of the nernst effect near
  quantum phase transitions in condensed matter and in dyonic black holes},\
  }\href {https://doi.org/10.1103/PhysRevB.76.144502} {\bibfield  {journal}
  {\bibinfo  {journal} {Phys. Rev. B}\ }\textbf {\bibinfo {volume} {76}},\
  \bibinfo {pages} {144502} (\bibinfo {year} {2007})}\BibitemShut {NoStop}%
\bibitem [{\citenamefont {Lucas}\ and\ \citenamefont
  {Sachdev}(2015)}]{LUCAS2015239}%
  \BibitemOpen
  \bibfield  {author} {\bibinfo {author} {\bibfnamefont {A.}~\bibnamefont
  {Lucas}}\ and\ \bibinfo {author} {\bibfnamefont {S.}~\bibnamefont
  {Sachdev}},\ }\bibfield  {title} {\bibinfo {title} {Conductivity of weakly
  disordered strange metals: From conformal to hyperscaling-violating
  regimes},\ }\href
  {https://doi.org/https://doi.org/10.1016/j.nuclphysb.2015.01.017} {\bibfield
  {journal} {\bibinfo  {journal} {Nuclear Physics B}\ }\textbf {\bibinfo
  {volume} {892}},\ \bibinfo {pages} {239} (\bibinfo {year}
  {2015})}\BibitemShut {NoStop}%
\bibitem [{\citenamefont {Lucas}\ \emph {et~al.}(2014)\citenamefont {Lucas},
  \citenamefont {Sachdev},\ and\ \citenamefont {Schalm}}]{PhysRevD.89.066018}%
  \BibitemOpen
  \bibfield  {author} {\bibinfo {author} {\bibfnamefont {A.}~\bibnamefont
  {Lucas}}, \bibinfo {author} {\bibfnamefont {S.}~\bibnamefont {Sachdev}},\
  and\ \bibinfo {author} {\bibfnamefont {K.}~\bibnamefont {Schalm}},\
  }\bibfield  {title} {\bibinfo {title} {Scale-invariant hyperscaling-violating
  holographic theories and the resistivity of strange metals with random-field
  disorder},\ }\href {https://doi.org/10.1103/PhysRevD.89.066018} {\bibfield
  {journal} {\bibinfo  {journal} {Phys. Rev. D}\ }\textbf {\bibinfo {volume}
  {89}},\ \bibinfo {pages} {066018} (\bibinfo {year} {2014})}\BibitemShut
  {NoStop}%
\bibitem [{\citenamefont {Hartnoll}\ \emph {et~al.}(2016)\citenamefont
  {Hartnoll}, \citenamefont {Lucas},\ and\ \citenamefont
  {Sachdev}}]{hartnoll2018holographic}%
  \BibitemOpen
  \bibfield  {author} {\bibinfo {author} {\bibfnamefont {S.~A.}\ \bibnamefont
  {Hartnoll}}, \bibinfo {author} {\bibfnamefont {A.}~\bibnamefont {Lucas}},\
  and\ \bibinfo {author} {\bibfnamefont {S.}~\bibnamefont {Sachdev}},\
  }\bibfield  {title} {\bibinfo {title} {{Holographic quantum matter}},\
  }\href@noop {} {\  (\bibinfo {year} {2016})},\ \Eprint
  {https://arxiv.org/abs/1612.07324} {arXiv:1612.07324 [hep-th]} \BibitemShut
  {NoStop}%
\bibitem [{\citenamefont {Mandal}\ and\ \citenamefont
  {Lucas}(2020)}]{ips_andy}%
  \BibitemOpen
  \bibfield  {author} {\bibinfo {author} {\bibfnamefont {I.}~\bibnamefont
  {Mandal}}\ and\ \bibinfo {author} {\bibfnamefont {A.}~\bibnamefont {Lucas}},\
  }\bibfield  {title} {\bibinfo {title} {Sign of viscous magnetoresistance in
  electron fluids},\ }\href {https://doi.org/10.1103/PhysRevB.101.045122}
  {\bibfield  {journal} {\bibinfo  {journal} {Phys. Rev. B}\ }\textbf {\bibinfo
  {volume} {101}},\ \bibinfo {pages} {045122} (\bibinfo {year}
  {2020})}\BibitemShut {NoStop}%
\bibitem [{\citenamefont {Sinner}\ and\ \citenamefont
  {Ziegler}(2012)}]{PhysRevB.86.155450}%
  \BibitemOpen
  \bibfield  {author} {\bibinfo {author} {\bibfnamefont {A.}~\bibnamefont
  {Sinner}}\ and\ \bibinfo {author} {\bibfnamefont {K.}~\bibnamefont
  {Ziegler}},\ }\bibfield  {title} {\bibinfo {title} {Renormalized transport
  properties of randomly gapped two-dimensional dirac fermions},\ }\href
  {https://doi.org/10.1103/PhysRevB.86.155450} {\bibfield  {journal} {\bibinfo
  {journal} {Phys. Rev. B}\ }\textbf {\bibinfo {volume} {86}},\ \bibinfo
  {pages} {155450} (\bibinfo {year} {2012})}\BibitemShut {NoStop}%
\bibitem [{\citenamefont {Ziegler}(2013)}]{Ziegler2013}%
  \BibitemOpen
  \bibfield  {author} {\bibinfo {author} {\bibfnamefont {K.}~\bibnamefont
  {Ziegler}},\ }\bibfield  {title} {\bibinfo {title} {Dynamical symmetry
  breaking in a 2d electron gas with a spectral node},\ }\href
  {https://doi.org/10.1140/epjb/e2013-40482-7} {\bibfield  {journal} {\bibinfo
  {journal} {The European Physical Journal B}\ }\textbf {\bibinfo {volume}
  {86}},\ \bibinfo {pages} {391} (\bibinfo {year} {2013})}\BibitemShut
  {NoStop}%
\bibitem [{\citenamefont {Ziegler}(2009)}]{PhysRevB.79.195424}%
  \BibitemOpen
  \bibfield  {author} {\bibinfo {author} {\bibfnamefont {K.}~\bibnamefont
  {Ziegler}},\ }\bibfield  {title} {\bibinfo {title} {Diffusion in the random
  gap model of monolayer and bilayer graphene},\ }\href
  {https://doi.org/10.1103/PhysRevB.79.195424} {\bibfield  {journal} {\bibinfo
  {journal} {Phys. Rev. B}\ }\textbf {\bibinfo {volume} {79}},\ \bibinfo
  {pages} {195424} (\bibinfo {year} {2009})}\BibitemShut {NoStop}%
\bibitem [{\citenamefont {Wegner}(1987)}]{WEGNER1987210}%
  \BibitemOpen
  \bibfield  {author} {\bibinfo {author} {\bibfnamefont {F.}~\bibnamefont
  {Wegner}},\ }\bibfield  {title} {\bibinfo {title} {Anomalous dimensions for
  the nonlinear sigma-model, in $2+\varepsilon$ dimensions (ii)},\ }\href
  {https://doi.org/https://doi.org/10.1016/0550-3213(87)90145-3} {\bibfield
  {journal} {\bibinfo  {journal} {Nuclear Physics B}\ }\textbf {\bibinfo
  {volume} {280}},\ \bibinfo {pages} {210} (\bibinfo {year}
  {1987})}\BibitemShut {NoStop}%
\bibitem [{\citenamefont {Ziegler}(2015)}]{Ziegler_2015}%
  \BibitemOpen
  \bibfield  {author} {\bibinfo {author} {\bibfnamefont {K.}~\bibnamefont
  {Ziegler}},\ }\bibfield  {title} {\bibinfo {title} {Quantum transport with
  strong scattering: beyond the nonlinear sigma model},\ }\href
  {https://doi.org/10.1088/1751-8113/48/5/055102} {\bibfield  {journal}
  {\bibinfo  {journal} {Journal of Physics A: Mathematical and Theoretical}\
  }\textbf {\bibinfo {volume} {48}},\ \bibinfo {pages} {055102} (\bibinfo
  {year} {2015})}\BibitemShut {NoStop}%
\bibitem [{\citenamefont {Negele}\ and\ \citenamefont
  {Orland}(1988)}]{zbMATH01727741}%
  \BibitemOpen
  \bibfield  {author} {\bibinfo {author} {\bibfnamefont {J.~W.}\ \bibnamefont
  {Negele}}\ and\ \bibinfo {author} {\bibfnamefont {H.}~\bibnamefont
  {Orland}},\ }\href@noop {} {\emph {\bibinfo {title} {Quantum many-particle
  systems}}},\ Vol.~\bibinfo {volume} {68}\ (\bibinfo  {publisher} {Redwood
  City, CA: Addison-Wesley Publishing},\ \bibinfo {year} {1988})\BibitemShut
  {NoStop}%
\bibitem [{\citenamefont {Ziegler}(1997)}]{PhysRevB.55.10661}%
  \BibitemOpen
  \bibfield  {author} {\bibinfo {author} {\bibfnamefont {K.}~\bibnamefont
  {Ziegler}},\ }\bibfield  {title} {\bibinfo {title} {Scaling behavior and
  universality near the quantum hall transition},\ }\href
  {https://doi.org/10.1103/PhysRevB.55.10661} {\bibfield  {journal} {\bibinfo
  {journal} {Phys. Rev. B}\ }\textbf {\bibinfo {volume} {55}},\ \bibinfo
  {pages} {10661} (\bibinfo {year} {1997})}\BibitemShut {NoStop}%
\bibitem [{\citenamefont {Forster}(1975)}]{Forster-HFBSCF}%
  \BibitemOpen
  \bibfield  {author} {\bibinfo {author} {\bibfnamefont {D.}~\bibnamefont
  {Forster}},\ }\href@noop {} {\emph {\bibinfo {title} {{Hydrodynamic
  Fluctuations, Broken Symmetry, and Correlation Functions}}}}\ (\bibinfo
  {publisher} {W. A. Benjamin},\ \bibinfo {address} {Reading},\ \bibinfo {year}
  {1975})\BibitemShut {NoStop}%
\bibitem [{\citenamefont {Freire}(2017)}]{Freire-AP_2017}%
  \BibitemOpen
  \bibfield  {author} {\bibinfo {author} {\bibfnamefont {H.}~\bibnamefont
  {Freire}},\ }\bibfield  {title} {\bibinfo {title} {{Memory matrix theory of
  the dc resistivity of a disordered antiferromagnetic metal with an effective
  composite operator}},\ }\href {https://doi.org/10.1016/j.aop.2017.07.001}
  {\bibfield  {journal} {\bibinfo  {journal} {Ann. Phys. (N. Y.)}\ }\textbf
  {\bibinfo {volume} {384}},\ \bibinfo {pages} {142} (\bibinfo {year}
  {2017})}\BibitemShut {NoStop}%
\bibitem [{\citenamefont {Mandal}\ and\ \citenamefont
  {Freire}(2021)}]{ips-freire1}%
  \BibitemOpen
  \bibfield  {author} {\bibinfo {author} {\bibfnamefont {I.}~\bibnamefont
  {Mandal}}\ and\ \bibinfo {author} {\bibfnamefont {H.}~\bibnamefont
  {Freire}},\ }\bibfield  {title} {\bibinfo {title} {Transport in the non-fermi
  liquid phase of isotropic luttinger semimetals},\ }\href
  {https://doi.org/10.1103/PhysRevB.103.195116} {\bibfield  {journal} {\bibinfo
   {journal} {Phys. Rev. B}\ }\textbf {\bibinfo {volume} {103}},\ \bibinfo
  {pages} {195116} (\bibinfo {year} {2021})}\BibitemShut {NoStop}%
\bibitem [{\citenamefont {Freire}\ and\ \citenamefont
  {Mandal}(2021)}]{ips-freire2}%
  \BibitemOpen
  \bibfield  {author} {\bibinfo {author} {\bibfnamefont {H.}~\bibnamefont
  {Freire}}\ and\ \bibinfo {author} {\bibfnamefont {I.}~\bibnamefont
  {Mandal}},\ }\bibfield  {title} {\bibinfo {title} {Thermoelectric and thermal
  properties of the weakly disordered non-fermi liquid phase of luttinger
  semimetals},\ }\href
  {https://doi.org/https://doi.org/10.1016/j.physleta.2021.127470} {\bibfield
  {journal} {\bibinfo  {journal} {Physics Letters A}\ }\textbf {\bibinfo
  {volume} {407}},\ \bibinfo {pages} {127470} (\bibinfo {year}
  {2021})}\BibitemShut {NoStop}%
\bibitem [{\citenamefont {Nandkishore}\ and\ \citenamefont
  {Parameswaran}(2017)}]{rahul-sid}%
  \BibitemOpen
  \bibfield  {author} {\bibinfo {author} {\bibfnamefont {R.~M.}\ \bibnamefont
  {Nandkishore}}\ and\ \bibinfo {author} {\bibfnamefont {S.~A.}\ \bibnamefont
  {Parameswaran}},\ }\bibfield  {title} {\bibinfo {title} {Disorder-driven
  destruction of a non-fermi liquid semimetal studied by renormalization group
  analysis},\ }\href {https://doi.org/10.1103/PhysRevB.95.205106} {\bibfield
  {journal} {\bibinfo  {journal} {Phys. Rev. B}\ }\textbf {\bibinfo {volume}
  {95}},\ \bibinfo {pages} {205106} (\bibinfo {year} {2017})}\BibitemShut
  {NoStop}%
\bibitem [{\citenamefont {{Mandal}}\ and\ \citenamefont
  {{Nandkishore}}(2018)}]{ips-rahul}%
  \BibitemOpen
  \bibfield  {author} {\bibinfo {author} {\bibfnamefont {I.}~\bibnamefont
  {{Mandal}}}\ and\ \bibinfo {author} {\bibfnamefont {R.~M.}\ \bibnamefont
  {{Nandkishore}}},\ }\bibfield  {title} {\bibinfo {title} {{Interplay of
  Coulomb interactions and disorder in three-dimensional quadratic band
  crossings without time-reversal symmetry and with unequal masses for
  conduction and valence bands}},\ }\href
  {https://doi.org/10.1103/PhysRevB.97.125121} {\bibfield  {journal} {\bibinfo
  {journal} {\prb}\ }\textbf {\bibinfo {volume} {97}},\ \bibinfo {eid} {125121}
  (\bibinfo {year} {2018})}\BibitemShut {NoStop}%
\bibitem [{\citenamefont {Mandal}\ and\ \citenamefont
  {Lee}(2015)}]{ips-uv-ir1}%
  \BibitemOpen
  \bibfield  {author} {\bibinfo {author} {\bibfnamefont {I.}~\bibnamefont
  {Mandal}}\ and\ \bibinfo {author} {\bibfnamefont {S.-S.}\ \bibnamefont
  {Lee}},\ }\bibfield  {title} {\bibinfo {title} {Ultraviolet/infrared mixing
  in non-fermi liquids},\ }\href {https://doi.org/10.1103/PhysRevB.92.035141}
  {\bibfield  {journal} {\bibinfo  {journal} {Phys. Rev. B}\ }\textbf {\bibinfo
  {volume} {92}},\ \bibinfo {pages} {035141} (\bibinfo {year}
  {2015})}\BibitemShut {NoStop}%
\bibitem [{\citenamefont {Mandal}(2016)}]{ips-uv-ir2}%
  \BibitemOpen
  \bibfield  {author} {\bibinfo {author} {\bibfnamefont {I.}~\bibnamefont
  {Mandal}},\ }\bibfield  {title} {\bibinfo {title} {{UV/IR Mixing In Non-Fermi
  Liquids: Higher-Loop Corrections In Different Energy Ranges}},\ }\href
  {https://doi.org/10.1140/epjb/e2016-70509-4} {\bibfield  {journal} {\bibinfo
  {journal} {Eur. Phys. J. B}\ }\textbf {\bibinfo {volume} {89}},\ \bibinfo
  {pages} {278} (\bibinfo {year} {2016})}\BibitemShut {NoStop}%
\bibitem [{\citenamefont {Eberlein}\ \emph {et~al.}(2016)\citenamefont
  {Eberlein}, \citenamefont {Mandal},\ and\ \citenamefont
  {Sachdev}}]{ips-subir}%
  \BibitemOpen
  \bibfield  {author} {\bibinfo {author} {\bibfnamefont {A.}~\bibnamefont
  {Eberlein}}, \bibinfo {author} {\bibfnamefont {I.}~\bibnamefont {Mandal}},\
  and\ \bibinfo {author} {\bibfnamefont {S.}~\bibnamefont {Sachdev}},\
  }\bibfield  {title} {\bibinfo {title} {Hyperscaling violation at the
  ising-nematic quantum critical point in two-dimensional metals},\ }\href
  {https://doi.org/10.1103/PhysRevB.94.045133} {\bibfield  {journal} {\bibinfo
  {journal} {Phys. Rev. B}\ }\textbf {\bibinfo {volume} {94}},\ \bibinfo
  {pages} {045133} (\bibinfo {year} {2016})}\BibitemShut {NoStop}%
\bibitem [{\citenamefont {Mandal}(2017)}]{ips-c2}%
  \BibitemOpen
  \bibfield  {author} {\bibinfo {author} {\bibfnamefont {I.}~\bibnamefont
  {Mandal}},\ }\bibfield  {title} {\bibinfo {title} {Scaling behaviour and
  superconducting instability in anisotropic non-fermi liquids},\ }\href
  {https://doi.org/https://doi.org/10.1016/j.aop.2016.11.009} {\bibfield
  {journal} {\bibinfo  {journal} {Annals of Physics}\ }\textbf {\bibinfo
  {volume} {376}},\ \bibinfo {pages} {89 } (\bibinfo {year}
  {2017})}\BibitemShut {NoStop}%
\bibitem [{\citenamefont {Pimenov}\ \emph {et~al.}(2018)\citenamefont
  {Pimenov}, \citenamefont {Mandal}, \citenamefont {Piazza},\ and\
  \citenamefont {Punk}}]{ips-fflo}%
  \BibitemOpen
  \bibfield  {author} {\bibinfo {author} {\bibfnamefont {D.}~\bibnamefont
  {Pimenov}}, \bibinfo {author} {\bibfnamefont {I.}~\bibnamefont {Mandal}},
  \bibinfo {author} {\bibfnamefont {F.}~\bibnamefont {Piazza}},\ and\ \bibinfo
  {author} {\bibfnamefont {M.}~\bibnamefont {Punk}},\ }\bibfield  {title}
  {\bibinfo {title} {Non-fermi liquid at the fflo quantum critical point},\
  }\href {https://doi.org/10.1103/PhysRevB.98.024510} {\bibfield  {journal}
  {\bibinfo  {journal} {Phys. Rev. B}\ }\textbf {\bibinfo {volume} {98}},\
  \bibinfo {pages} {024510} (\bibinfo {year} {2018})}\BibitemShut {NoStop}%
\bibitem [{\citenamefont {Mandal}(2020)}]{ips-nfl-u1}%
  \BibitemOpen
  \bibfield  {author} {\bibinfo {author} {\bibfnamefont {I.}~\bibnamefont
  {Mandal}},\ }\bibfield  {title} {\bibinfo {title} {Critical fermi surfaces in
  generic dimensions arising from transverse gauge field interactions},\ }\href
  {https://doi.org/10.1103/PhysRevResearch.2.043277} {\bibfield  {journal}
  {\bibinfo  {journal} {Phys. Rev. Research}\ }\textbf {\bibinfo {volume}
  {2}},\ \bibinfo {pages} {043277} (\bibinfo {year} {2020})}\BibitemShut
  {NoStop}%
\end{thebibliography}%

\appendix

\section{Example of a particle-hole symmetric tight-binding model}
\label{app:example}

As an example for a system with PH symmetry, we consider the tight-binding Hamiltonian on the honeycomb lattice with nearest neighbor hopping. The honeycomb lattice is bipartite and consists of two triangular sublattices. The nearest-neighbor hopping connects these two sublattices, such
that we can write the hopping Hamiltonian in the sublattice representation as
\[
H=\begin{pmatrix}
0 & h \\
h^T & 0 
\end{pmatrix}
\ ,
\]
where the hopping term $h^T$ is the transpose of $h$. $H$ can be expanded in terms of the Pauli matrices as
\[
H={\cal H}_1\,\sigma^1+{\cal H}_2\,\sigma^2
\ ,\ \ 
{\cal H}_1= \left (h+h^T \right )/2
\ ,\ \ 
{\cal H}_2=i \left (h-h^T \right )/2
\ .\]
Thus, we get $H\to \sigma^3 \,H\,\sigma^3=-H$, i.e., $S=\sigma^3$ in Sec.~\ref{sect:rpa}, as the PH transformation. PH-symmetric disorder can be implemented as corrugations or random strain in the hopping terms.

\section{Calculation details of the perturbation theory}
\label{app:details}

The vertex in Eq.~(\ref{vertex}) reads
\[
V_{\br_1\br_2\br_3\br_4}
=\sum \limits_{j_1,...,j_4=1,2}
\left (1-\delta_{\br_1j_1,\br_3j_3} \right )
h_{\br_1 j_1,\br_2j_2} \,h^\dagger_{\br_2j_2,\br_3j_3}
h_{\br_3j_3,\br_4j_4} \,h^\dagger_{\br_4j_4,\br_1j_1}
\ ,
\]
which can be decomposed with the help of trace terms as 
\begin{align}
& \left (1-\delta_{\br_1\br_3} \right )
Tr_2 \left (h_{\br_1\br_2}\,h^\dagger_{\br_2\br_3}
\,h_{\br_3\br_4}\,h^\dagger_{\br_4\br_1} \right ) 
+ \delta_{\br_1\br_3}
\left [Tr_2
\left (h_{\br_1\br_2} \,h^\dagger_{\br_2\br_3}
\,h_{\br_3\br_4} \,h^\dagger_{\br_4\br_1} \right )
-\sum \limits_j
\left (h_{\br_1\br_2} \,h^\dagger_{\br_2\br_1} \right )_{jj}
\left (h_{\br_1\br_4} \,h^\dagger_{\br_4\br_1} \right )_{jj} \right ]
\nonumber \\ & =Tr_2
\left (h_{\br_1\br_2}h^\dagger_{\br_2\br_3}\,h_{\br_3\br_4}
\,h^\dagger_{\br_4\br_1} \right )
-\delta_{\br_1\br_3}\sum \limits_j
\left (h_{\br_1\br_2} \,h^\dagger_{\br_2\br_1} \right)_{jj}
\left (h_{\br_1\br_4}\,h^\dagger_{\br_4\br_1} \right)_{jj}\ .
\end{align}
Since we get $h\to S\,h\, S^{-1}=h^\dagger$ from a PH transformation, the first term obeys the relation
\beq
Tr_2\left (h_{\br_1\br_2}\,h^\dagger_{\br_2\br_3}
\,h_{\br_3\br_4}h^\dagger_{\br_4\br_1} \right)
=Tr_2\left (h^\dagger_{\br_1\br_2}\,h_{\br_2\br_3}
\,h^\dagger_{\br_3\br_4} \,h_{\br_4\br_1}
 \right ) .
\eeq
Moreover, $\left \langle C\right \rangle_\alpha^{-1}$ is real symmetric, as mentioned in Sec.~\ref{sect:rpa}.
With these two properties, the perturbation expansion up to two loops can be rewritten
as
\begin{align}
& \frac{1}{Z_0}\int_\Psi \bar{\Psi}_\br\,\Psi_{\br'}
\left \langle e^{-\Psi\cdot C \, \bar{\Psi} }\right \rangle_\alpha
\approx
G_{\br\br'}
\left [ 1+V_{\br_1\br_2\br_3\br_4} \left (G_{\br_1\br_2} \,G_{\br_3\br_4}
-G_{\br_1\br_4} \,G_{\br_3\br_2} \right ) \right ] \,.
\end{align}
Noting that
\begin{align}
 V_{\br_1\br_2\br_3\br_4} \left [
G_{\br\br_1} \left (G_{\br_2\br_3}\,G_{\br_4\br'}-G_{\br_2\br'} \, G_{\br_4\br_3} \right )
+G_{\br\br_3}
\left (G_{\br_2\br'} \,G_{\br_4\br_1}-G_{\br_2\br_1}\, G_{\br_4\br'} \right )
\right ] = G_{\br\br'}\,,
\end{align}
since the vertex is invariant under  $V_{\br_1\br_2\br_3\br_4}
\to V_{\br_4\br_1\br_2\br_3}$ (cyclic permutation of its indices). We also have taken into account the appropriate signs, which reflect the fermionic statistics of the field $\Psi$. These properties finally lead to the result
\beq
\frac{1}{Z_0}\int_\Psi \left \langle
e^{ -\Psi\cdot C \,\bar{\Psi}}\right \rangle_\alpha
\approx
1+V_{\br_1\br_2\br_3\br_4}
\left (G_{\br_1\br_2} \,G_{\br_3\br_4}-G_{\br_1\br_4} \,G_{\br_3\br_2} \right )
= 1 \ .
\eeq

\end{document}